\documentclass{iau379}

\usepackage{graphicx}
\usepackage{amsmath}
\usepackage{multirow}

\begin{document}

\lefttitle{B\'ilek et al.}
\righttitle{IAU Symposium 379: Template}

\jnlPage{1}{7}
\jnlDoiYr{2023}
\doival{10.1017/xxxxx}

\def\pk#1 {\textcolor{blue}{#1} } 

\aopheadtitle{Proceedings of IAU Symposium 379}
\editors{P. Bonifacio,  M.-R. Cioni, F. Hammer, M. Pawlowski, and S. Taibi, eds.}

\title{Do old globular clusters in low-mass galaxies disprove modified gravity?}

\author{B\'ilek M.$^{1,2,3}$, Zhao H.$^4$, Famaey B.$^3$, Nagesh S. T. $^3$, Combes F.$^{1,2}$, M\"uller O.$^5$, Hilker M.$^6$, Kroupa P.$^{7,8}$, Ibata R.$^3$}
\affiliation{$^1$ LERMA, Observatoire de Paris, CNRS, PSL Univ., Sorbonne Univ., 75014 Paris, France\\
$^2$ Coll\`ege de France, 11 place Marcelin Berthelot, 75005 Paris, France\\
$^3$ Universit\'e de Strasbourg, CNRS, Observatoire astronomique de Strasbourg, UMR 7550, F-67000 Strasbourg, France\\ 
$^4$ Scottish Universities Physics Alliance, University of St Andrews, North Haugh, St Andrews, Fife KY16 9SS, UK  \\
$^5$ Institute of Physics, Laboratory of Astrophysics, Ecole Polytechnique F\`ed\`erale de Lausanne (EPFL), 1290 Sauverny, Switzerland\\
$^6$ European Southern Observatory, Karl-Schwarzschild-Strasse 2, 85748 Garching bei M\"unchen, Germany
$^7$Helmholtz-Institut f\"ur Strahlen- und Kernphysik (HISKP), Universit\"at Bonn, Nu{\ss}allee 14-16, D-53115 Bonn, Germany \\
$^8$Astronomical Institute, Faculty of Mathematics and Physics, Charles University in Prague, V Hole{\v s}ovi{\v c}k{\' a}ch 2, CZ-18000 Praha, Czech Republic
}

\begin{abstract}
The controversy ``dark matter vs. modified gravity'' constitutes a major topic of discussion. It was proposed that dynamical friction could be used to discriminate between the two alternatives. Analytic calculations indicate that, with modified gravity, globular clusters (GCs) of low-mass galaxies experience much stronger dynamical friction than in the equivalent system with Newtonian gravity and dark matter. As a result, in modified gravity the old GCs of low mass galaxies should have already settled in the centers of the galaxies. This is not observed. Here we report on our efforts to verify the analytic results by self-consistent simulations with the MOND-type (modified Newtonian dynamics) gravity. The core stalling mechanism, that was not considered in the analytic calculations, prevents GCs to settle in centers of ultra-diffuse galaxies. For isolated dwarf galaxies, which are gas-rich objects, supernova explosions prevent the GCs from settling.
\end{abstract}

\begin{keywords}
Dwarf galaxies, globular clusters, dynamical friction, modified gravity
\end{keywords}

\maketitle

\section{Introduction}

The nature of gravity and inertia is not clear yet. A generally accepted theory of quantum gravity is missing. There are indications that our imperfect knowledge of the laws or gravity or inertia is the reason why we encounter the missing mass problem. The strongest indication for this is the fact that dynamical properties of most galaxies can be predicted from the distribution of their visible material \citep{mcgaugh16,lelli17,bil19,milg19} using the prescriptions of modified Newtonian dynamics (MOND, \citealp{milg83a}), a  paradigm of modified gravity and/or inertia. This indicates that if indeed gravity or inertia need to be updated, then the correct theory has to follow MOND in certain regimes. Nevertheless, alternatives to such theories as the explanation of the missing mass problem exist, such as the hypothesis of dark matter. 

One can test these hypotheses using dynamical friction \citep{Kroupa2015}. It is a force acting against the direction of motion of two interacting $N$-body systems. It happens because the relative orbital energy of the systems is transferred into the internal energy of their constituents, that is stars, gas, or dark matter particles. 

We are still rather at the beginning of the investigation of dynamical friction in MOND. Its exact functioning depends to an unknown degree on the particular MOND theory. We thus consider hereafter only the MOND modified gravity theories of \citet{BM84} and \citet{qumond}. Simulations have shown that once we deal with interactions of two objects with similar masses (such as a major merger of galaxies), dynamical friction turns out to be weaker in MOND than in an equivalent Newtonian system with collisionless dark matter  \citep{ciotti04}. On the other hand, analytic calculations indicate that once the two interacting objects have a large mass ratio, dynamical friction is stronger in the MOND system than in the equivalent Newtonian system. The enhancement of dynamical friction is generally larger for objects with a lower surface brightness. According to analytic calculations, the strong dynamical friction makes  globular clusters (GCs) of low-mass galaxies spiral down the center of their host on the timescale of a few Gyr. This opposes the observed ages of the GCs in such galaxies of many Gyr. We decided to verify the analytic results by high-resolution simulations.

\section{Case of isolated ultra-diffuse galaxies}
Ultra-diffuse galaxies (UDGs) are characterized by masses of dwarf galaxies and sizes of giant galaxies. It was found that not only they contain old GCs, but these GCs can be exceptionally massive, exceeding $10^6\,M_\odot$. This  should make the settling of the GCs in the center of their host galaxy (i.e. the process known as ``sinking'') even faster. We simulated the evolution of such objects in MOND in  \citet{bil21}.

We used the MOND version of the adaptive-mesh-refinement code RAMSES \citep{ramses,por}. The galaxy was simulated having an effective radius of 2\,kpc and stellar mass of $2\times10^8\,M_\odot$, i.e. resembling the UDG NGC\,1052-DF2 that hosts very massive GCs. The galaxy was modeled as a spherical $N$-body isotropic system, the GCs as single massive particles. The maximum spatial resolution was 50\,pc and the particles of the galaxy had $20\,M_\odot$ each. The galaxy was simulated as isolated.

We first let a GC having $10^6\,M_\odot$ fall on the UDG from  5\,kpc with a zero relative velocity. The GC first experienced strong damping of its oscillation by dynamical friction. Nevertheless, the orbital decay of the GC suddenly slowed down once the GC moved within the inner 1\,kpc of the galaxy. It is unlikely that the change was caused by an insufficient resolution: the friction became less efficient when the size of the orbit was about ten times larger than the resolution and the behavior did not change after increasing the resolution. The phenomenon is known from simulations with Newtonian gravity too under the name of core stalling. It happens in galaxies having nearly harmonic gravitational potentials ($\phi(r)\propto r^2$) near their centers. Dynamical friction becomes ineffective because stars have equal orbital periods in such a potential. This behavior is not captured by the analytic calculations that predicted the fast sinking of GCs of low-mass galaxies in MOND. This is because of the simplifying assumptions done during their derivation.

With Newtonian gravity, the Chandrasekhar formula is usually used for evaluating the strength of dynamical friction. Its analog for MOND has been proposed by \citet{sanchezsalcedo06}. This formula indicated the problem of fast sinking of GCs of low-mass galaxies in MOND. We were tested the validity of this formula.  We repeated the simulation of the free-fall of a GC varying the mass of the GC. Then we explored the motion of a $10^6\,M_\odot$ GC for different orbital shapes. It turned out that the S\'anchez-Salcedo formula works well, unless the GC moves inside the inner 1\,kpc of the galaxy, where the core stalling occurs. This is the same result as for the Chandrasekhar formula in Newtonian gravity. We found that the value of Coulomb logarithm in the S\'anchez-Salcedo formula varies qualitatively with the orbital shape of the GCs as in the Chandrasekhar formula \citep{chan97}. 

We then simulated an evolution of a system of ten GCs. The GCs were distributed in the simulated galaxy in a similar fashion as observed in the DF2 galaxy. We considered two types of GC mass functions -- the standard one, observe for most galaxies, and the mass function seen in the galaxy DF2 that is biased toward high GC masses. The system was evolved for a Hubble time. It turned out that the outer GCs are little affected by dynamical friction. Only the inner ones approached toward the galaxy center, but a full sinking was disabled by core stalling. In the result, the GC system became more centrally concentrated. This evolution was stronger for the DF2-like GC mass function. We predicted the profiles of surface number-density that should be observed in isolated UDGs. It should be noted however, that the DF2-like GC mass function has been reported so far only for the galaxy DF2 itself, for which our simulations do not apply because it is not isolated -- it is exposed to a strong external field effect from a neighboring massive elliptical \citep{famaey18,haghi19}. For such a galaxy, the Newtonian simulations of \citet{dutta20} are more relevant.

We then utilized symmetries of the equations governing the motion of the GCs to scale our simulation for UDGs of different masses and sizes. It helped us to verify for nearly all observed UDGs that the presence of GCs in them does not exclude MOND. Finally, we found that GCs can evaporate through tidal heating during their encounters.

  \begin{figure}
    \centering\includegraphics[width = 0.7\textwidth]{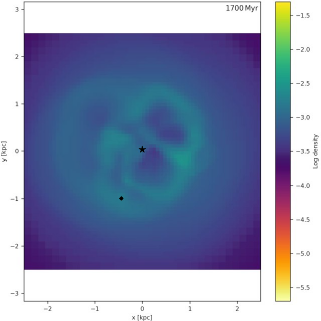}
    \caption{Example snapshot from a simulation with supernovae. The star indicates the barycenter of the galaxy, the dot indicates the GC.}
    \label{fig:sn}
  \end{figure}

\section{Case of isolated dwarf galaxies}
In an ongoing project, we study the survival of GCs of isolated dwarf galaxies. The galaxies are gas-rich rotating objects. Equations indicate the GCs of the least massive dwarfs sink the fastest. We reviewed the literature on observations of GCs of isolated dwarfs and found that the least massive isolated dwarfs that are known to host GC candidates have baryonic masses of about $2\times10^7\,M_\odot$, and that GCs are common in galaxies over about $10^8\,M_\odot$.

We simulated the motion of GCs in a galaxy of the mass of $10^8\,M_\odot$. The dwarf was initiated with 90\% of gas, representing the likely state 10\,Gyr ago, when old GCs are formed. The scale length of both the gas and stellar disks were 2\,kpc. The GCs were again simulated as a point masses. The grid was refined if the mass in a cell exceeded $4\times10^4\,M_\odot$ with the limit of 0.5\,pc.

We first explored the motion of a GC of a relatively typical mass of $10^5\,M_\odot$ in simulations without star formation and supernova feedback. If the GC initially moves inside the plane of the disk, either on a circular co-rotating obit or a radial orbit, the GC approaches the center of the galaxy in about 1\,Gyr. For a counter-rotating, polar or axial orbit, the GC does not sink within 10\,Gyr.  A GC of $10^4\,M_\odot$ does not show substantial sinking on any orbit within 10\,Gyr.

\begin{figure}
    \centering\includegraphics[width = 0.75\textwidth]{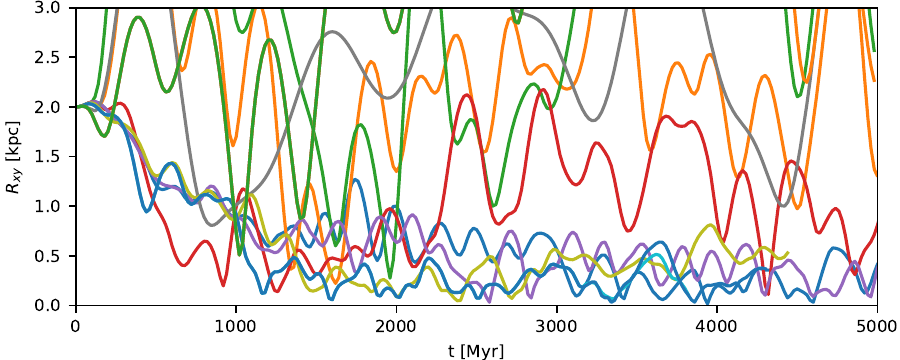}
    \caption{Evolution of the distance between the GC and galaxy barycenter. The different lines indicate the results for the different prescriptions for star formation and supernova feedback, such as the star formation timescale, efficiency or supernova kinetic energy fraction. }
    \label{fig:kicks}
  \end{figure}

We then included star formation and supernova feedback. We explored a wide range of star formation efficiencies and of parameters of the supernova models included in RAMSES. All simulations showed the same: supernovae prevent the GC from sinking regardless of the initial direction of motion of the GC.  In MOND, where most of the mass of the dwarf is in gas, supernova explosions lead to large fluctuations of gravitational potential (Fig.~\ref{fig:sn}). The fluctuations of the gravitational field give random kicks to the GC, which prevent it from settling at the galaxy center (Fig.~\ref{fig:kicks}). Supernova explosions can even change the sense of rotation of the GC around the center of the galaxy. With Newtonian gravity, this mechanism would be much reduced since dark matter, the dominant component, does not directly react to the supernovae.

\bibliographystyle{iaulike}

\end{document}